# Astrophysical *S* factor for the radiative capture $^{12}$N($p, \gamma$)$^{13}$O determined from the $^{14}$N($^{12}$N,$^{13}$O)$^{13}$C proton transfer reaction

A. Banu[*], T. Al-Abdullah[†], C. Fu[‡], C. A. Gagliardi, M. McCleskey, A. M. Mukhamedzhanov, G. Tabacaru[§], L. Trache, R. E. Tribble, Y. Zhai[**]

*Cyclotron Institute, Texas A&M University, College Station, Texas 77843, U.S.A.*

F. Carstoiu

*National Institute of Physics and Nuclear Engineering "Horia Hulubei", R-76900 Bucharest-Magurele, Romania*

V. Burjan and V. Kroha

*Institute of Nuclear Physics, Czech Academy of Sciences, CZ-250 68 Prague-Řež, Czech Republic*

The cross section of the radiative proton capture reaction on the drip line nucleus $^{12}$N was investigated using the Asymptotic Normalization Coefficient (ANC) method. We have used the $^{14}$N($^{12}$N,$^{13}$O)$^{13}$C proton transfer reaction at 12 MeV/nucleon to extract the ANC for $^{13}$O → $^{12}$N + $p$ and calculate from it the direct component of the astrophysical *S* factor of the $^{12}$N($p, \gamma$)$^{13}$O reaction. The optical potentials used and the DWBA analysis of the proton transfer reaction are discussed. For the entrance channel, the optical potential was inferred from an elastic scattering measurement carried out at the same time with the transfer measurement. From the transfer, we determined the square of the ANC, $C^2_{p_{1/2}}(^{13}O_{g.s.}) = 2.53 \pm 0.30$ fm$^{-1}$, and hence a value of 0.33(4) keV·b was obtained for the direct astrophysical *S* factor at zero energy. Constructive interference at low energies between the direct and resonant captures leads to an enhancement of $S_{total}(0) = 0.42(5)$ keV·b. The $^{12}$N($p, \gamma$)$^{13}$O reaction was investigated in relation to the evolution of hydrogen-rich massive Population III stars, for the role that it may play in the *hot pp*-chain nuclear burning processes, possibly occurring in such objects.

---

[*] banu@comp.tamu.edu
[†] Present address: Hashemite University, Zaqra, Jordan
[‡] Present address: National Institute of Standards and Technology (NIST), Gaithersburg, MD, U.S.A.
[§] On leave from National Institute of Physics and Nuclear Engineering (IFIN-HH), Bucharest, Romania
[**] Present address: University of Texas Health Science Center, San Antonio, TX, U.S.A.

## I. INTRODUCTION

Modern cosmology estimates that at the end of the cosmic dark ages, about 300 million years after the Big Bang, the first luminous objects in the universe, the so-called Population III stars, formed. Understanding the properties of the first stars and what impact they had in driving early cosmic evolution of the universe are key problems in modern cosmology. Supernova explosions that ended the lives of some of the first stars are responsible for the initial enrichment of the intergalactic medium with heavy chemical elements and consequently they had important effects on subsequent galaxy evolution. The most fundamental question about the Population III stars is how massive they typically were. Numerical simulations of the collapse of primordial H/He gas indicate that the first stars were predominantly very massive with masses larger than hundreds of solar masses [1]. Currently we do not have direct observational constraints on any of the properties of the first stars because not a single metal-free star has ever been detected. Even if such behemoth "zero metallicity" stars were formed, the astronomers may never find evidence of them due to their short lifetimes.

In 1986, Fuller et al. [2] addressed the classic problem of the evolution of supermassive stars (Population III non-rotating stars with masses greater than $10^5$ solar masses): given that such an object has formed and quasi-statically contracted to the point of dynamical instability, is the nuclear energy generated in the subsequent collapse enough to blow up the star? They modeled two possible scenarios—explosion or collapse—and concluded that non-rotating supermassive stars with zero metallicity will never explode but collapse into black holes. The smallest metallicity needed for an explosion was $Z = 0.005$. For the failed explosion, it was reasoned that in the short time scales of the collapse insufficient amounts of $^{12}$C and other heavy elements are produced by the triple alpha process, $3\alpha \rightarrow$ $^{12}$C. By $t = 10^4$ s into the collapse, the authors found that the central temperatures and densities are of the order of $10^9$ K and 1000 g/cm$^3$, respectively. Under these conditions hydrogen is burned rapidly by the *rp*-process. According to Ref. [2], an explosion can not occur for two reasons: first, while the star dynamically collapses waiting for the $3\alpha$ process to generate enough catalytic nuclei to burn hydrogen in the hot CNO cycle and by the *rp*-process, it builds up a huge infall kinetic energy that can not be overcome by nuclear energy generation; second, as the temperature rises to near $10^9$ K, electron-positron pairs are substantially produced and so neutrinos, with the result that all the extra thermal gas pressure created by the *rp*-process goes into neutrino energy losses.

Three years later in 1989, Wiescher et al. [3] proposed alternative ways to bypass the triple alpha process and produce CNO material—the *hot pp* chains and *rap*-processes:

*ppIV:* $^7$Be($p,\gamma$)$^8$B($p,\gamma$)$^9$C($\beta^+\nu$)$^9$B($p$)
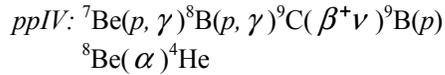
$^8$Be($\alpha$)$^4$He

*pp-V:* $^7$Be($\alpha,\gamma$)$^{11}$C($\beta^+\nu$)$^{11}$B($p,2\alpha$)$^4$He

*rap-I:* $^7$Be($p,\gamma$)$^8$B($p,\gamma$)$^9$C($\alpha,p$)$^{12}$**N**($p,\gamma$)$^{13}$**O**
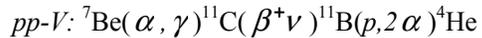
($\beta^+\nu$)$^{13}$N($p,\gamma$)$^{14}$O

*rap-II:* $^7$Be($\alpha,\gamma$)$^{11}$C($p,\gamma$)$^{12}$**N**($p,\gamma$)$^{13}$**O**
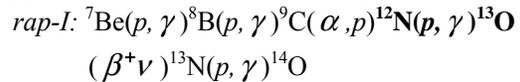
($\beta^+\nu$)$^{13}$N($p,\gamma$)$^{14}$O

*rap-III:* $^7$Be($\alpha,\gamma$)$^{11}$C($p,\gamma$)$^{12}$N($\beta^+\nu$)$^{12}$C($p,\gamma$)
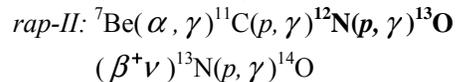
$^{13}$N($p,\gamma$)$^{14}$O

*rap-IV:* $^7$Be($\alpha,\gamma$)$^{11}$C($\alpha,p$)$^{14}$N($p,\gamma$)$^{15}$O.
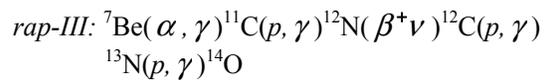
The $^{12}$N($p,\gamma$)$^{13}$O reaction is an important branching point in the *rap*-I and *rap*-II chains.

The outcome of Wiescher et al.'s study is that for densities in excess of 1 g/cm$^3$, temperatures of $\sim 3 \times 10^8$ K are sufficient to process material from the *pp*-chains to CNO nuclei. The question to be answered is whether the *rap*-processes can lead to the production of sufficient amounts of CNO material that may also result in an explosion of supermassive objects with lower metallicity than that proposed by Fuller et al.'s study. Taking into account that only primordial abundances of $^2$H and $^3$He are initially available to produce CNO nuclei through the *hot pp-chain* and *rap-process* scenario, the authors of Ref. [3] came up with a mass fraction of CNO material of $\sim 3\times 10^{-4}$, an order of magnitude smaller than the threshold of $5\times 10^{-3}$ found in Ref. [2] as a trigger for an explosion of non-rotating supermassive stars. However, the energy release by the *rap*-processes was not taken into consideration in the original hydrodynamical calculations of the collapse. Wiescher *et al.* suggested that this energy release may moderate or even halt the collapse. Hence, the inversion of the collapse may be possible at much smaller CNO amounts.

More recent simulation studies in massive metal-free stars between 120 and 1000 solar masses indicate that a metallicity as small as $\sim 1\times 10^{-9}$ is sufficient to stop the contraction and supply energy through the hot CNO cycle for hydrogen burning [4].

Additional nuclear data are needed in order to put the scenario based on $^{12}$C formation via *rap*-processes on a firmer quantitative basis. The new experimental information on the $^{12}$N($p,\gamma$)$^{13}$O reaction reported here, along with input for other reactions in dedicated full nuclear reaction network calculations, might play an important role in modelling of the evolution and nucleosynthesis in those lower mass Population III stars.

In the following we report on the determination of the astrophysical *S* factor and the reaction rate for the radiative proton capture $^{12}$N($p,\gamma$)$^{13}$O from the study of the $^{14}$N($^{12}$N,$^{13}$O)$^{13}$C peripheral transfer reaction using the Asymptotic Normalization Coefficient (ANC) method [5]. In section II the experimental procedure and the setup are described. The data analysis of the elastic scattering measurement, from which the optical potential of the entrance channel was inferred, is presented in section III, and the optical model potential parameters needed in the Distorted Wave Born Approximation (DWBA) analysis of the transfer reaction are discussed. In sections IV and V, the discussion is focused on the analysis of the proton transfer data (IV) and on the determination of the astrophysical *S* factor, as well as on the rate for the $^{12}$N($p,\gamma$)$^{13}$O reaction and its implications in stellar environments (V).

## II. EXPERIMENT

The measurement was performed at the Texas A&M University Cyclotron Institute with a radioactive beam of $^{12}$N. The $^{12}$C primary beam was delivered by the K500 superconducting cyclotron with an intensity of 150 pnA and impinged on a cryogenic H$_2$ gas cell that was operated at a pressure of 2.2 atm. Its entrance and exit windows were made of $13\mu$m- and $4\mu$m-thick Havar foils, respectively. The gas cell was cooled with liquid nitrogen to obtain higher density at a lower pressure, thus increasing the yield of the radioactive beam while minimizing the thickness of windows needed. Due to a large negative Q-value (-18.12 MeV) of the (*p,n*) reaction used in inverse kinematics to produce the $^{12}$N secondary beam, the energy of the primary beam had to be large, 23 MeV/nucleon, resulting in a $^{12}$N beam energy larger than the typical energy regime of 10-12 MeV/nucleon where we have a tested procedure [6] to obtain optical model potentials needed in the DWBA analysis of peripheral transfer reactions. To bring it down to 12 MeV/nucleon, the energy of the secondary beam was degraded by a $250\mu$m-thick Al foil placed behind the gas cell. The resulting $^{12}$N beam separated by the Momentum Achromat Recoil Spectrometer (MARS) [7] had a purity of around 99.8% and a rate of around $2\times 10^5$ pps. It impinged on a composite melamine target (C$_3$H$_6$N$_6$) located at the final focal plane of MARS, and of a thickness of $1.58\pm 0.05$ mg/cm$^2$ (measured offline with a $^{228}$Th alpha source). The production and separation of the secondary beam in MARS was done with a procedure similar to the one described in Ref. [8]. With momentum defining slits in MARS open to $\pm 1.0$ cm, we had a beam energy spread of ±1.2 % around the mean value of 139 MeV. The $^{12}$N beam was tuned at the location of the target using a 1 mm-thick, 16-strip position sensitive detector, with the primary beam intensity attenuated by about $10^3$. The energy of the $^{12}$N projectiles in the middle of the target was 137.6 MeV and the beam size at the secondary target position was measured to be 3.5 mm × 4.0 mm FWHM (horizontal × vertical). The last pair of slits in MARS removed any impurities in the beam that had a charge-to-mass ratio different from that of the fully stripped $^{12}$N. Another pair of slits (dubbed SL3), located just after the last quadrupoles of MARS at 75 cm upstream from the secondary target, were used to define the beam angular spread. Two different settings were used during the experiment: one setting with the SL3 slits opened wider (4.2 cm × 2 cm) to maximize the intensity of the beam for the measurement of the proton transfer reaction and for large-angle elastic scattering, and another setting with the SL3 slits narrower (2.2 cm × 2 cm) to improve the beam angular definition, used only for the elastic scattering measurement at forward angles.

We used an experimental detection setup that served for reaction channel selection and particle scattering angle determination. It consisted of four modular $\Delta E$-$E$ telescopes (5 × 5 cm$^2$ in area), placed 182 mm downstream from the melamine target. One pair of telescopes (1 and 2) was positioned symmetrically up-down at ±13 mm from the beam axis with an angular coverage of 4°- 19° (in the laboratory frame), while the second pair (telescopes 3 and 4) was positioned symmetrically left-right at ±52 mm covering angles from 16° to 30°. The same detection system was used to study the proton transfer reaction ($^7$Be,$^8$B) to obtain information about $^7$Be($p, \gamma$)$^8$B and is shown in Fig. 1 of Ref. [9].

The front $\Delta E$ detectors were 16-strip position-sensitive silicon detectors, 110 $\mu$m thick (telescopes 1 and 2) and 65 $\mu$m thick (telescopes 3 and 4). The back $E$-residual detectors had the same area coverage and were all 500 $\mu$m thick. We recorded the signals from one end of each resistive strip as well as the total energy loss from the back (ohmic side) of the $\Delta E$ detector and the particle residual energy detected in the $E$ detector. The position along a strip was determined from the strip charge signal and the total energy (back signal) in the $\Delta E$ detectors. A position calibration was done using four masks with five 0.8-mm-wide slots 8 mm apart attached to the front side of each telescope, from which the position resolution along the strips was inferred to be 0.4 mm (FWHM). The corresponding detector position resolution in the perpendicular direction was given by the width of the strips (3.1 mm).

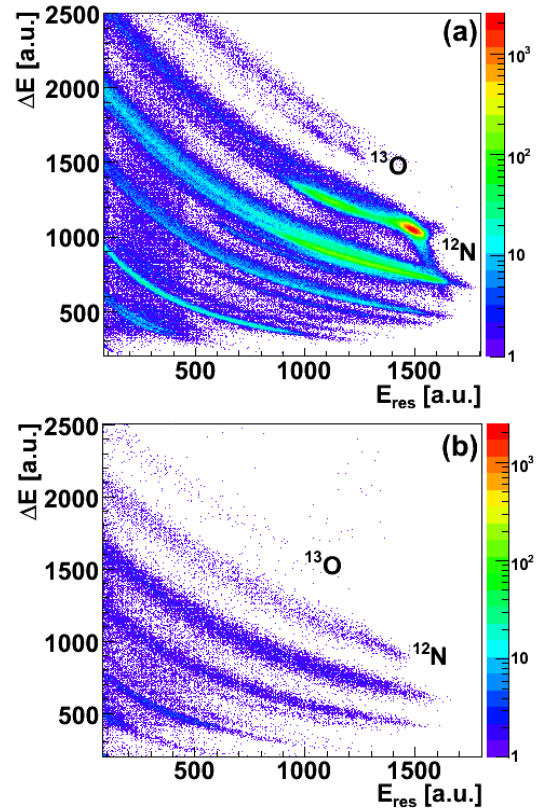

Fig 1. (Color online) Particle identification plots $\Delta E$ (vertical) vs. $E$ residual (horizontal) for (a) telescope 1, covering the small angle region, and (b) telescope 3 at large angles.

Figure 1 illustrates $\Delta E$-$E$ particle identification plots at small angles (Fig. 1 (a)) and at large angles (Fig. 1 (b)). While at small angles the cross sections for both the elastic and transfer channels are large and the $^{12}$N and $^{13}$O loci are clearly visible, at large angles there is a reasonable yield only for the elastic channel. The energy resolutions were around 2 MeV (FWHM) in both $\Delta E$ and $E$ detectors, sufficient to provide good particle identification in all four telescopes.

The beam normalization was provided by counting the secondary beam ions in a plastic scintillator detector coupled to a photomultiplier tube which was placed at 0° downstream of the target and the Si-detector array. A wire-mesh screen with a "transparency" determined to be $11.3 \pm 0.4\%$ was used to reduce the rate of secondary beam particles giving signals in the scintillator detector.

## III. ELASTIC SCATTERING AND OPTICAL MODEL POTENTIALS

A complication in the data analysis was given by the composite nature of the melamine target, $C_3H_6N_6$. With a software gate on the $^{12}$N locus in Fig. 1 (a), a two-dimensional plot, kinetic energy vs. scattering angle, was produced to disentangle the scattering off the three different species of nuclei in the melamine target. While the scattering off $^1$H nuclei was easily identified, we could not distinguish between scattering off $^{12}$C and $^{14}$N nuclei, except at the very largest angles. Therefore, for consistency we treated them together for the whole angular range of the measurements with the melamine target.

Considering the elastic scattering off the melamine target as it would have happened with respect to $^{14}$N nuclei only, we have reconstructed the corresponding Q-value shown in Fig. 2, which has an energy resolution of 2 MeV (FWHM). Here the events are selected from the $\Delta E$-$E$ plot of Fig. 1 (a) that represent the $^{12}$N locus. The elastic peak is centered around 0 MeV and a software Q-value cut from -2.5 MeV to 2.5 MeV was applied to select the elastic channel corresponding to the scattering off

$^{12}$C and $^{14}$N nuclei. The small left-side inelastic peak corresponds to the -4.43 MeV energy of the first excited state in $^{12}$C. No bound excited states exist in $^{12}$N.

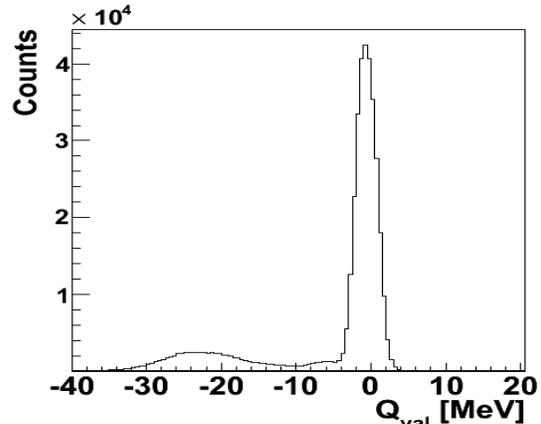

Fig. 2: Reconstructed Q-value for the $^{12}$N elastic scattering channel off the melamine target detected at small angles. The distribution around zero energy corresponds to elastic scattering on both $^{12}$C and $^{14}$N nuclei in the target. The small bump centered around -6 MeV (includes the kinematics shift) corresponds to the inelastic channel $^{12}$N-$^{12}$C$^*$, while the left-most distribution is the elastic scattering off $^1$H nuclei in the target.

Excited states in $^{14}$N have energies of 2.31 MeV, 3.95 MeV, 5.11 MeV and 5.83 MeV, among which only the first excited state would be problematic for the aforementioned Q-value cut of the elastic channel. However, the transition between the ground state of $^{14}$N and its first excited state is a pure spin-flip *M1* transition, which is unlikely to occur in inelastic scattering. Indeed, the inelastic excitation of this state was found to be very weak in a previous high resolution study of $^{13}$C($^{14}$N,$^{14}$N)$^{13}$C at a similar energy [10] and, therefore, we neglected it here. The Q-value of the elastic scattering was also reconstructed with kinematics as for scattering off C-nuclei in the melamine target, making the Q-value selection safe against the 4.43 MeV first excited state of $^{12}$C. Similar results were obtained in this case for the experimental yields of the elastic angular distributions.

For the DWBA analysis of the transfer reaction of interest, $^{14}$N($^{12}$N,$^{13}$O)$^{13}$C, reliable Optical Model Potentials (OMPs) for both the entrance channel ($^{12}$N-$^{14}$N) and exit channel ($^{13}$O-$^{13}$C) are needed to calculate the corresponding incoming/outgoing distorted scattering wave functions. In our experiment, we measured simultaneously the proton transfer reaction and the elastic scattering of $^{12}$N ions off the melamine (C$_3$H$_6$N$_6$) target, which enabled us to extract the OMP for the entrance channel from the analysis of the elastic data. However, in the part of the measurement where we needed to maximize the secondary beam intensity to obtain good transfer data (slits SL3 open), the angular resolution of the beam precluded a clear observation of the Fraunhofer oscillations in the elastic scattering angular distribution. Therefore, for part of the measurement we closed the SL3 slits at the expense of beam intensity, reducing the angular spread of the beam to 0.8°, which was sufficient to preserve the Fraunhofer oscillations.

For the elastic data analysis, we have used semi-microscopic double-folding optical potentials. The procedure was established from a systematic search [6] of optical potentials for use in the description of elastic and transfer reactions involving stable and loosely bound $p$-shell nuclei. It has also been proved to work fairly well for the elastic scattering of radioactive nuclei such as $^{7}$Be, $^{8}$B [9], $^{11}$C, $^{13}$N [11] and $^{17}$F [12]. The folding model uses the effective nucleon-nucleon interaction from the nuclear matter approach of Jeukenne, Lejeune, and Mahaux (JLM) [13] with parameters tuned by Bauge *et al.* [14] for nucleon-nucleus scattering.

It was established in Ref. [6] that the potentials calculated with the double-folding procedure need to be renormalized. In that global analysis of elastic data, the double-folding potential

$$U_{DF}(r) = N_V V(r, t_V) + i N_W W(r, t_W)$$

is a four-parameter potential with renormalization coefficients $N_V$, $N_W$ and the range parameters $t_V$, $t_W$. Good results were obtained with fixed values for the range parameters $t_V = 1.2$ fm and $t_W = 1.75$ fm and only $N_V$ and $N_W$ kept free. The authors found that while the depth of the real potential needs a substantial renormalization (on average: $N_V = 0.37(1)$), the imaginary part needs no such renormalization ($N_W = 1.0(1)$).

The $p$-shell nucleus $^{12}$N with a proton separation energy $S_p = 600$ keV is a loosely-bound nucleus. The angular distributions corresponding to its elastic scattering off $^{14}$N and $^{12}$C were each calculated separately (in the center-of-mass frame) using the respective $^{14}$N+$^{12}$N and $^{12}$C+$^{12}$N double-folding potentials computed with the same values for the range parameters $t_{V(W)}$ and renormalized with the same coefficients $N_{V(W)}$ (the validity of this assumption is discussed at the end of the section). After that they were transformed into the laboratory frame. In Fig. 3, the two elastic scattering components are summed in the laboratory frame taking into account the stoichiometry of carbon and nitrogen in melamine. The normalization is chosen such that we plot the quantity:

$$\frac{d\sigma}{d\Omega_{lab}}(mel) = \frac{d\sigma}{d\Omega_{lab}}(^{14}N) + 0.5 \frac{d\sigma}{d\Omega_{lab}}(^{12}C).$$

The experimental elastic angular distribution, binned in steps of 0.5° (in the laboratory frame) is plotted in Fig. 3 in comparison to the calculated cross sections. Here the data points are plotted with their statistical errors only.

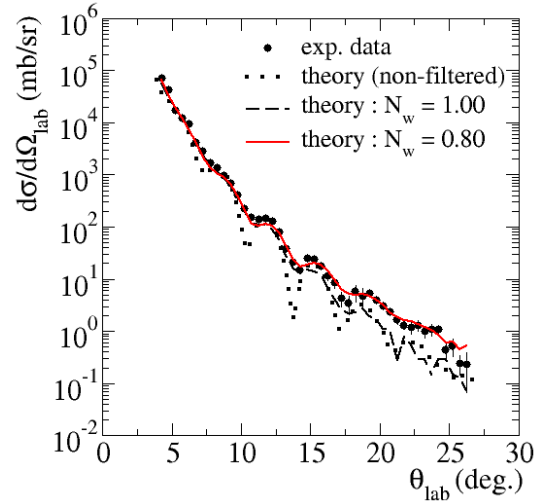

Fig. 3. (Color online) Angular distribution for elastic scattering of $^{12}$N off $^{14}$N and $^{12}$C nuclei in the melamine target. The theoretical calculation (dotted curve) was filtered with the experimental conditions through a Monte Carlo (MC) simulation to obtain the dashed curve. The calculations were carried out with $N_V = 0.37$, $N_W = 1.0$, $t_V = 1.2$ fm, $t_W = 1.75$ fm. The *red solid* curve was calculated with $N_W = 0.80$.

The dotted curve represents the double-folding potential calculation with the aforementioned OMP parameters of Ref. [6]: $N_V = 0.37$, $N_W = 1.0$, $t_V = 1.2$ fm, $t_W = 1.75$ fm. The dashed curve shows the same distribution filtered through a Monte Carlo (MC) simulation that accounts for the real experimental conditions. The MC simulation included the position, size, and divergence of the beam on the target, the calculated angular distributions, and the finite resolution of the detectors.

From the comparison it is clear that the main features are reproduced using the "standard" parameters, in particular the position of the minima and maxima (attesting a good real part of the potential), but they reproduce poorly the elastic data at larger angles, indicating that the absorptive potential is too strong. There are two ways to remedy this: either decrease the standard renormalization parameter $N_W$ of the imaginary depth of the double-folding potential without changing the imaginary range parameter, or keep the standard value of the renormalization parameter as $N_W = 1.0$ and change instead the imaginary range parameter $t_W$. We found that for a renormalization $N_W = 0.85$-$0.80$ the elastic data is well matched, while an overestimation of the elastic cross section occurs for $N_W = 0.75$. Such a renormalization of the imaginary part slightly different from unity for the *p*-shell nucleus $^{12}$N is similar with what was found for the *sd*-shell nucleus $^{17}$F [12]: $N_V = 0.63$, $N_W = 0.90$, $t_V = 1.2$ fm, $t_W = 1.75$ fm ($^{17}$F + $^{14}$N). Keeping the renormalization parameter $N_W = 1.0$, an overall reasonable fit to the data was also found for a range parameter $t_W = 1.2$ fm.

The three solutions mentioned above are summarized in Table 1. In Fig 3 (*solid red curve*), we plotted as an example the case for $N_W = 0.80$, after binning and convolution with the experimental resolutions through MC simulations. No renormalization of the absolute values of the experimental elastic cross sections was needed. Because they gave equivalent description of the elastic data, we adopted all three solutions for the DWBA analysis of the transfer channel of interest.

Table I. The double-folding optical-model parameters and the corresponding $\chi^2$ per degree of freedom for the calculations compared to the data of $^{12}$N elastic scattering off the melamine target.

| OMP | $N_V$ | $N_W$ | $t_v$(fm) | $t_w$(fm) | $\chi^2/N$ |
|---|---|---|---|---|---|
| (1) | 0.37 | 0.85 | 1.2 | 1.75 | 38.7 |
| (2) | 0.37 | 0.80 | 1.2 | 1.75 | 38.2 |
| (3) | 0.37 | 1.0 | 1.2 | 1.2 | 43.1 |

In order to check the validity of using the same renormalization and range parameters in treating the elastic scattering off $^{14}$N and $^{12}$C nuclei in the melamine target, we carried out a separate experiment to measure the elastic scattering of $^{12}$N projectiles off of a $^{12}$C target. The results are shown in Fig. 4 where the measured elastic angular distribution is compared to a double-folding potential calculation with OMPs corresponding to set (3) in Table I.

The experimental data energy resolution did not allow us to disentangle the inelastic scattering of the first excited $2^+$ state in $^{12}$C from the ground state scattering corresponding to the elastic channel. This was related to the fact that the carbon target thickness was approximately 7 mg/cm$^2$. As a result we had a more pronounced smearing of the Fraunhofer oscillations in the angular distribution. The inelastic contribution was calculated using the coupled-channel code ECIS [15] with a deformation parameter $\beta_{coul} = 0.582$ taken from the literature [16] (assuming the same deformation length for the nuclear component, $\beta_{coul}R_c = \beta_{nucl}R_n$). The calculations for both the elastic and inelastic channels were convoluted in Fig. 4 with the experimental resolutions.

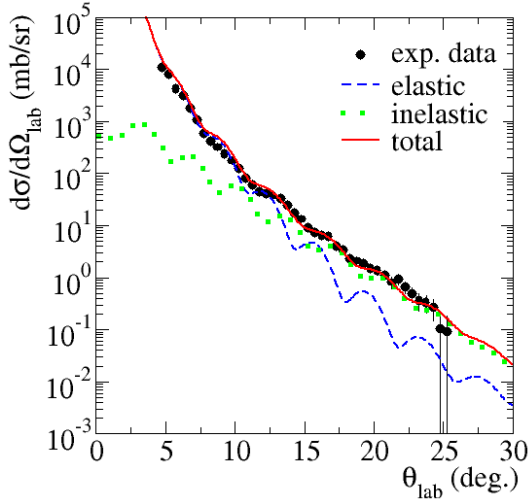

Fig. 4: (Color online) Comparison between experimental data and theoretical calculation that describes the scattering off $^{12}$C target. (plotted in the laboratory frame as per Fig. 3 for comparison). Shown here in *blue* is a double-folding potential calculation of the elastic channel. The *green* line represents coupled-channel calculation of the inelastic scattering corresponding to the first excited $2^+$ state in $^{12}$C, and the *red* line is the incoherent sum between these two scattering components. See text for details.

As Fig. 4 illustrates, the theoretical calculation (red curve) reproduces well the experimental data. We can thus conclude that the double-folding potential renormalization and range parameters used for the composite melamine target give a good description of the elastic data on the carbon target.

Moreover, the current results for the 12 MeV/nucleon elastic scattering data in the *p*-shell region are in agreement with the previous works [6], enabling us to assess the validity of the double-folding procedure based on the JLM effective interaction to predict optical model potentials for the use in DWBA calculations. Therefore, lacking measured elastic data for the exit channel $^{13}$O-$^{13}$C of the transfer reaction under investigation here, we have similarly computed the corresponding double-folding potential and assumed the same renormalization and range parameters as extracted here for the entrance channel.

In addition to the analysis of the $^{12}$N elastic scattering data using double-folding potentials, we have attempted analyses using phenomenological potentials with standard Wood-Saxon shapes. The best results were obtained with volume plus surface potential terms [17]. Because this parametrization involves a large number of free parameters (12), we used it with caution. The second attempt was using volume terms only (6 free parameters). Two sets of Woods-Saxon potentials were found with continuous ambiguities within each class. Both shallow potentials (with volume integrals $J_V \approx 70$ MeV·fm$^3$, $J_W \approx 40$ MeV·fm$^3$) and deep potentials (with $J_V \approx 240$ MeV·fm$^3$, $J_W \approx 160$ MeV·fm$^3$) gave equivalent reasonable fits to the elastic scattering data. We also have used these potentials to assess the dependence of the DWBA calculations for the transfer on the optical potentials used.

## IV. ANALYSIS OF THE PROTON TRANSFER DATA

Radiative proton capture reactions at stellar energies are peripheral processes due to the presence of the Coulomb barrier and occur with very small cross sections. We can therefore study such reactions employing indirect methods. This is the case for the radiative proton capture reaction $^{12}$N$(p,\gamma)^{13}$O studied here via the ANC method using the $^{14}$N$(^{12}$N,$^{13}$O$)^{13}$C proton transfer reaction. At 12 MeV/nucleon, the transfer process is peripheral with the advantage that it happens at energies above the Coulomb barrier, thereby yielding a much larger cross section than the deeply sub-Coulomb radiative proton capture at astrophysically relevant energies.

The basis of the application of the ANC method for $^{12}$N$(p,\gamma)^{13}$O entails the fact that the cross section for this peripheral reaction (note the small proton binding energy in $^{13}$O, $\varepsilon_p = 1.515$ MeV) is completely determined by the ANC for $^{13}$O $\rightarrow$ $^{12}$N + $p$. This ANC can be extracted indirectly from the peripheral proton transfer reaction by normalizing the calculated DWBA cross sections to the experimental transfer data, provided that the ANC for the other vertex of the reaction is known.

In our particular case, a proton from the $^{14}$N nuclei in the melamine target, occupying either the $1p_{1/2}$ or $1p_{3/2}$ orbitals, is transferred most probably to the $1p_{1/2}$ orbital in the $^{13}$O nucleus. The following expression is obtained for the experimental differential cross section in the DWBA analysis:

$$\sigma_{exp} = \left(C^{13O}_{p_{1/2}}\right)^2 \left\{ \left(\frac{C^{14N}_{p_{1/2}}}{b^{13O}_{p_{1/2}} b^{14N}_{p_{1/2}}}\right)^2 \sigma^{DW}_{p_{\frac{1}{2}} p_{\frac{1}{2}}} + \left(\frac{C^{14N}_{p_{3/2}}}{b^{13O}_{p_{1/2}} b^{14N}_{p_{3/2}}}\right)^2 \sigma^{DW}_{p_{\frac{1}{2}} p_{\frac{3}{2}}} \right\}$$

where $lj$ are the usual quantum numbers that characterize in this case the proton single orbitals involved, $C_{lj}$ are the ANCs, $b_{lj}$ are the single-particle ANCs of the normalized single-particle wave functions, and $\sigma^{DW}_{lj}$ are DWBA cross sections. The ANCs for the vertex $^{14}$N → $^{13}$C + p were determined from previous studies [10, 18].

In the following we discuss the determination of the ANC of interest, $C_{p_{1/2}}(^{13}O)$. In Fig. 5 the reconstructed Q-value for the transfer reaction is shown (obtained with an initial $^{13}$O cut on the two-dimensional plot $\Delta E$-$E$ in Fig. 1 (a)). The transfer channel of interest, $^{14}$N($^{12}$N,$^{13}$O)$^{13}$C, was selected as the peak on the right-hand side, whereas the peak on the left-hand side corresponds to the reaction $^{12}$C($^{12}$N,$^{13}$O)$^{11}$B on the $^{12}$C nuclei in the melamine target.

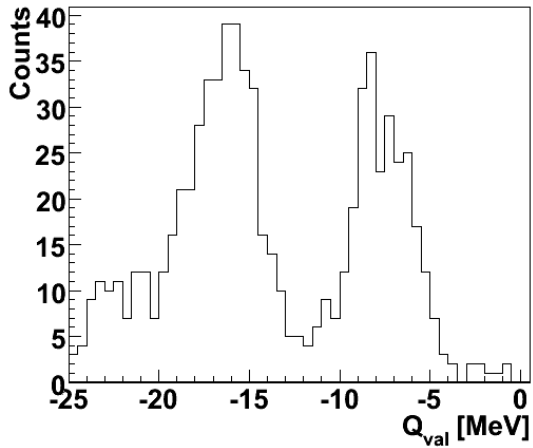

Fig. 5: Reconstructed Q-value of the ($^{12}$N,$^{13}$O) proton transfer reaction employing the melamine target. The peak on the right-hand side corresponds to the transfer channel of interest, $^{14}$N($^{12}$N,$^{13}$O)$^{13}$C, whereas the other peak corresponds to the transfer channel $^{12}$C($^{12}$N,$^{13}$O)$^{11}$B.

The corresponding experimental angular distribution is plotted in Fig. 6 in the centre-of-mass frame. The solid curve is a DWBA fit for the proton transfer calculation carried out with the finite-range DWBA code PTOLEMY [19].

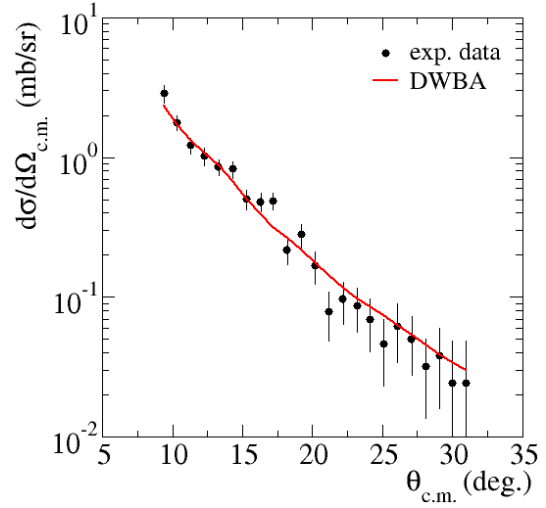

Fig. 6: (Color online) Transfer reaction angular distribution for $^{14}$N($^{12}$N,$^{13}$O)$^{13}$C. The *solid red* curve is the fit with DWBA calculation.

The distorted waves were calculated using the three sets of double-folding optical potentials presented in the previous section. A Woods-Saxon nuclear potential was used to bind the transferred proton in the $^{13}$O nucleus, characterized by the reduced radius and diffuseness ($r$, $a$), the Coulomb radius $r_C$ = 1.2 fm, and the spin-orbit $V_{SO}$ = 18.6 MeV of Ref. [20]. The depth of the bound state potential was adjusted to reproduce the experimental proton binding energy in $^{13}$O, and was found to be $V$ = 43.47 MeV.

In Fig. 7 we compare the ground state spectroscopic factor $S_{p_{1/2}}$ and the squared ANC, $C^2_{p_{1/2}}(^{13}O)$, extracted for the geometrical parameters of the proton binding potential ranging from $r$ = 1.0-1.3 fm and $a$ = 0.5-0.7 fm (varying in 0.1 fm steps), as functions of the corresponding single particle ANC[††]. As the figure clearly illustrates, the spectroscopic factor depends strongly on the choice of the single-particle potential parameters, while the ANC squared varies by less than 9 % over the full range. If we exclude the geometrical parameters that give unreasonable sizes of $^{13}$O like ($r = 1.0$ fm, $a = 0.5$ fm), (1.1 fm, 0.5 fm), (1.2 fm, 0.7 fm) and (1.3 fm, 0.7 fm), the variation of the $C^2_{p_{1/2}}$ is less than 5%, whereas the corresponding spectroscopic factor varies over 24%. This shows the peripherality of the transfer reaction.

---

[††] The DWBA analysis was done here using the double-folding potential (3) in Table I.

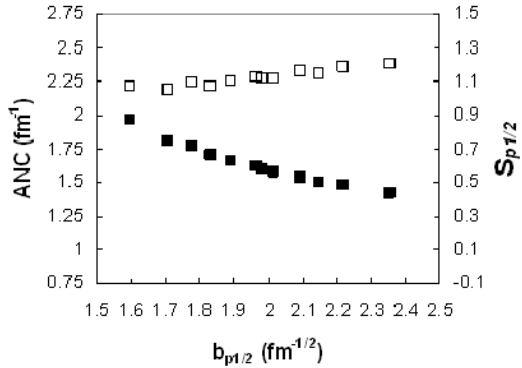

Fig. 7: The comparison between the spectroscopic factor $S_{p_{1/2}}$ (full squares) and the $C^2_{p_{1/2}}$ (open squares) extracted for the ground state of $^{13}$O as a function of the single particle ANC, $b_{p_{1/2}}$. See text for details.

Another indication of the peripheral character of the reaction is the localization of the transfer strength calculated with the DWBA code. The transition matrix elements for transfer peak around nucleus-nucleus relative orbital momentum values of 29-30, which correspond semi-classically to about $R = 6.26$ fm, to be compared with the grazing distance of $R_1+R_2 = 5.16$ fm between $^{12}$N projectiles and $^{14}$N target nuclei.

In Table II, we present the extracted values of $C^2_{p_{1/2}}(^{13}O)$ for each of the three sets of optical potentials discussed in the elastic scattering analysis. The results are reported for a proton binding potential with $r = 1.25$ fm, $a = 0.65$ fm, leading to the single-particle ANC, $b_{p_{1/2}} = 2.18$ fm$^{-1/2}$.

Table II. Extracted ANC values for the three double-folding potentials presented in Table I.

| OMP | $\chi^2_{elastic}/N$ | $C^2_{p_{1/2}}(^{13}O)$ [fm$^{-1}$] |
|---|---|---|
| (1) | 38.7 | 2.74 |
| (2) | 38.2 | 2.49 |
| (3) | 43.1 | 2.33 |

We have also carried out DWBA calculations using one of the phenomenological potentials found to describe well the elastic scattering data, as mentioned at the end of the previous section. We have used the Woods-Saxon potential with real and imaginary volume terms: $V = 160$ MeV, $r_V = 0.70$ fm, $a_V = 0.90$ fm, $W = 70.06$ MeV, $r_W = 0.85$ fm, $a_W = 0.86$ fm. We found a variation of only 3.5 % in the ANC relative to the value extracted using the double-folding potential (2) of Table I.

In Table III the contributions to the uncertainties in $C^2_{p_{1/2}}(^{13}O)$ are presented. Note that the total uncertainties are dominated by the choice of the optical model potentials in the DWBA analysis, while the systematic errors are beam normalization errors due primarily to the measurement of the target thickness, MC simulation and the estimation of the $^{12}$N beam intensity reduction in the plastic scintillator.

Table III: Contributions to the uncertainties in $C^2_{p_{1/2}}(^{13}O)$ determination.

| | |
|---|---|
| Statistical errors and fit | 4.6 % |
| Measurement systematic errors | 5.3 % |
| DWBA calculation systematic errors | |
| (a) Proton binding potential | 2.5 % |
| (b) OMP parameters | 8.0 % |
| (c) ANC for $^{14}$N | 4.9 % |
| Total | 11.9% |

Finally, we adopted an average of the extracted three values, weighted with the chi-square of the elastic data, yielding the value $C^2_{p_{1/2}}(^{13}O) = 2.53 \pm 0.30$ fm$^{-1}$. The corresponding spectroscopic factor $S_{p_{1/2}}(^{13}O_{g.s.}) = 0.53 \pm 0.06$, found for $r = 1.25$ fm and $a = 0.65$ fm, is in excellent agreement with the value of 0.537 extracted

## V. ASTROPHYSICAL *S* FACTOR AND REACTION RATE FOR $^{12}$N$(p,\gamma)^{13}$O

The asymptotic normalization coefficient, which is the amplitude of the tail of the projection of the bound state wave function of $^{13}$O on the two-body channel $^{12}$N + *p*, determines the overall normalization of the direct radiative capture astrophysical *S* factor for $^{12}$N$(p, \gamma)^{13}$O [22].

The first estimate of the $^{12}$N$(p, \gamma)^{13}$O reaction rate and of its astrophysical *S* factor was done in Ref. [3], where it was assumed that the reaction proceeds as an *E1* direct capture to the ground state of $^{13}$O, $J^{\pi} = 3/2^{-}$, and through a resonance at an excitation energy of $E_x$ = 2.75 MeV, $J^{\pi} = (3/2^{+})$ with its subsequent *E1* decay to the ground state. The radiative width of that resonance was suggested in Ref. [3] to have a value of $\Gamma_{\gamma}$ = 24 meV with one order of magnitude uncertainty, coming from a Weisskopf estimate of the transition strength. Recent work by B. B. Skorodumov et al. [23] measured the excitation function for resonance elastic scattering of *p* + $^{12}$N. The data were analyzed in the framework of the **R**-matrix formalism. The spin and parity $J^{\pi} = 1/2^{+}$ were found for the first excited state of $^{13}$O at an excitation energy of 2.69 (5) MeV. A resonance width $\Gamma$ = 0.45 (10) MeV was also determined.

In the following, we discuss the calculation of the direct and resonant captures for $^{12}$N$(p,\gamma)^{13}$O, and the interference between these two components. The calculations were performed using the **R**-matrix formalism (in L-S coupling) presented in brief in Ref. [23] and at large in Ref. [11]. In this formalism the radiative width amplitude is given by the sum of the internal (radial integral taken over the nuclear interior) and the external (outside the nuclear interior) matrix elements describing the radiative proton capture. In the case under consideration, the *E1* decay of the resonance to the ground state is the non-spin-flip transition $(l_i = 0, I = 1/2, J_i^{\pi} = 1/2^{+}) \rightarrow (l_f = 1, I = 1/2, J_f^{\pi} = 3/2^{-})$, where $l_i, J_i^{\pi}(l_f, J_f^{\pi})$ are the $p-^{12}$N relative orbital and total angular momenta for the initial-continuum and (final-bound) states of the radiative capture process, and *I* is the channel spin. The ANC of the overlap function of the ground states of $^{13}$O and $^{12}$N entering the external matrix element is $C_{l=1 I=1/2} = 2/3 C_{p_{1/2}}(^{13}O)$. We found that the external amplitude gives dominant contribution compared to the internal part, estimated in the single-particle approach.

Because the experimentally determined here $C^2_{p_{1/2}}(^{13}O) = 2.53 \pm 0.30$ fm$^{-1}$ is lower than $C^2_{p_{1/2}}(^{13}O) = 3.42^{\ddagger\ddagger}$ used in Ref. [23], and because in Ref. [23] the ANC determining the normalization of the external part of the radiative width amplitude was $C_{p_{1/2}}$ rather than $C_{l=1 I=1/2}$, the radiative width of the resonance calculated here $\Gamma_{\gamma}$ = 0.95 eV (for a channel radius *R* = 4.25 fm) is lower than the value of $\Gamma_{\gamma} \approx$ 3 eV obtained in Ref. [23], but significantly larger than the value of 24 meV used in Ref. [3].

It is interesting to note that the ANC method allows one to determine the low limit of the radiative width. We remind that the external part of the radiative width amplitude is complex, $f + ig$, because the resonant wave function in the external region is described by the outgoing scattering wave, while the internal part, $h$, is real. Hence, from $\Gamma_{\gamma} = \sqrt{(h+f)^2 + g^2} \geq g$ we calculated,

---

$^{\ddagger\ddagger}$ In the absence of experimental data for the ANC of $^{13}$O $\rightarrow$ $^{12}$N + *p*, Ref. [22] has calculated it from $C^2_{p_{1/2}} = S_{p_{1/2}} b^2_{p_{1/2}}$ with $b_{p_{1/2}} = 2.14$ fm$^{-1/2}$ (calculated with geometrical parameters of the proton bound state potential in $^{13}$O as $r_0$ =1.20 fm and *a* = 0.65 fm), and assuming $S_{p_{1/2}} = 0.75$.

model independent, the low limit of the radiative width to be g = 40 meV.

The direct capture amplitude is given by the sum of $I = 1/2$ and $3/2$ components. The first component interferes with the resonant amplitude. Within the same **R**-matrix formalism, due to the dominance of the external matrix element in the radiative width amplitude, we have found that the interference pattern of the resonant and direct capture amplitudes is constructive at energies below the resonance energy.

The total astrophysical $S$ factor was calculated for the coherent sum of the non-resonant (direct) and the resonant capture $I = 1/2$ amplitudes and non-coherent direct $I = 3/2$ component. We also took into account the non-coherent contribution from direct capture component with $l_i = 2$. The results are plotted in Fig. 8.

To determine the uncertainty of the calculated total astrophysical factor, we varied the channel radius from $R = 4.0$ fm to $R = 4.5$ fm, and obtained $\Gamma_\gamma = 0.9$ eV for $R = 4.0$ fm, and $\Gamma_\gamma = 1.1$ eV for $R = 4.5$ fm. We found the uncertainty in the total $S$ factor is 12 %, determined primarily by the uncertainty of the ANC.

Thus, taking into account the interference between the direct and the resonant captures of the $^{12}$N$(p,\gamma)^{13}$O reaction, we obtained for the total astrophysical factor $S$ factor at zero energy a value $S(0)$ = 0.42(5) keV·b, with the direct component dominating and contributing a value of 0.33(4) keV·b.

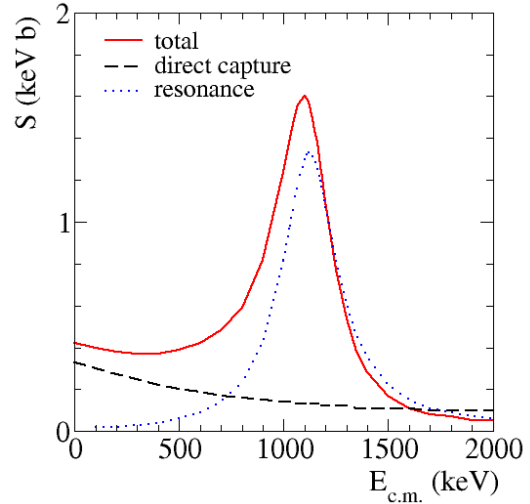

Fig. 8: (Color online) Astrophysical $S$ factor of the $^{12}$N$(p,\gamma)^{13}$O reaction as a function of the energy in the center-of-mass reference system. The dashed curve shows the direct capture component of the $S$ factor, while the dotted curve is the resonant component. The solid curve is the total astrophysical $S$ factor.

The astrophysical $S$ factor corresponding to the direct radiative capture $^{12}$N$(p,\gamma)^{13}$O was estimated in Ref. [3] to have an average value of $S_{DC} \approx 40$ keV·b, two orders of magnitude larger than the value found here. This difference has tremendous consequences for the reaction rate of the radiative capture under discussion.

Because the resonant state in $^{13}$O is broad, we had to use the full expression for reaction rates [24] and numerically integrate it to determine the total reaction rate for the radiative capture $^{12}$N$(p,\gamma)^{13}$O:

$$N_A \langle \sigma v \rangle_{total} = N_A \left(\frac{8}{\pi\mu}\right)^{1/2} \frac{1}{(kT)^{3/2}}$$

$$\times \left( \int_0^\infty S(E) \exp\left[ -\frac{E}{kT} - \left(\frac{E_G}{E}\right)^{1/2} \right] dE \right).$$

Here $N_A$ is Avogadro's number, $\mu$ represents the reduced mass of the system, $E$ is the energy in the center-of-mass, $S(E)$ is the total astrophysical $S$ factor, and $E_G$ is the Gamow energy (not to be confused with the maximum of the Gamow peak, see [24]). Given the variation of the astrophysical $S$ factor with energy in Fig. 8, we found sufficient to carry the numerical integration over energy up to an upper limit of 2000 keV. The result for the total reaction rate is plotted in Fig. 9 as a function of $T_9$, the temperature in units of $10^9$ K. The reaction rate calculation for the direct capture component with a potential model RADCAP [25] gave a consistent reaction rate evaluation. For comparison, in Fig 10 we plot the total reaction rate as evaluated in this work and the reaction rate for the direct capture only calculated as indicated in Ref. [3]. There is a difference of at least a factor of 60 between the two reaction rate evaluations along the temperature range considered.

Because of the large discrepancy in the values obtained for the astrophysical $S$ factor of $^{12}$N$(p,\gamma)^{13}$O by this work versus Ref. [3], the effective burning conditions at which the radiative capture of interest may take place need to be revised.

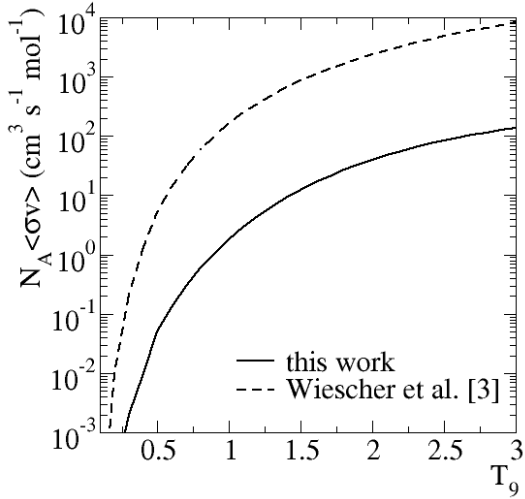

Fig. 9: Total reaction rate of the radiative capture $^{12}$N$(p,\gamma)^{13}$O determined by this work and plotted in comparison with the reaction rate corresponding to the direct capture only as evaluated in Ref. [3].

Our results are illustrated in Fig. 10. The solid curve indicates the conditions where the radiative capture reaction is of equal strength with the competing temperature- and density-independent $\beta^+$-decay of $^{12}$N with a half-life of 11 ms. Hence, the corresponding density values are determined from the Saha equation as

$$\rho_\beta^{(p,\gamma)} = \frac{\ln 2}{T_{1/2}(^{12}N)Y(^1H)N_A\langle\sigma v\rangle_{(p,\gamma)}} \text{ g cm}^{-3},$$

where $Y(H)$ is the hydrogen abundance equal to a value of 0.8. The proton capture will dominate over the beta decay above this equilibrium curve. In this region, the *rap*-I and *rap*-II processes will proceed through $^{12}$N$(p,\gamma)^{13}$O$(\beta^+\nu)^{13}$N$(p,\gamma)^{14}$O. Below the solid curve in Fig. 10, $^{12}$N $\beta^+$-decay will prevail leading to $^{12}$C, from where the conventional hot CNO cycle will take over the hydrogen burning.

In addition to its $\beta^+$-decay, $^{13}$O may also be depleted at sufficiently high temperature and density conditions by photodisintegration. The photodisintegration rate ($s^{-1}$) is given in Ref. [26] as

$$\lambda_{(\gamma,p)} = 9.87\cdot 10^9\, \omega^{-1}\left(\frac{A_1 A_2}{A_3}\right)^{3/2} T_9^{3/2} N_A\langle\sigma v\rangle_{(p,\gamma)}$$
$$\times \exp(-11.605 Q_6/T_9),$$

where $Q_6 = Q$ in MeV (Q-value), $A_1 = 12$, $A_2 = 1$, $A_3 = 13$, and $\omega = (2J_3+1)/(2J_1+1)(2J_2+1)$ is the statistical factor. Equating the rate expressions for the proton radiative capture on $^{12}$N and the photodisintegration of the resulting $^{13}$O, we end up with an expression for the density at which the two reactions compete with equal strength. This is illustrated in Fig. 10 by the dotted line. Note here that the *rap*-II process will dominate in the density-temperature region to the left of the dotted curve.

Another competing process responsible for $^{13}$O depletion may be the $^{13}$O$(\alpha,p)^{16}$F reaction. The $^{16}$F produced is proton-unbound and will decay immediately by proton emission to the ground state of $^{15}$O. Alpha capture on $^{14}$O as well as on $^{15}$O may link the *hot* pp chains [3] with the *rp*-process via the *rap*-processes.

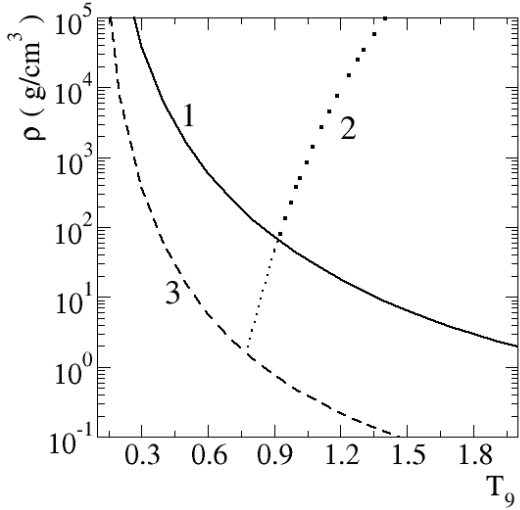

Fig. 10: Temperature and density conditions at which the $^{12}$N$(p,\gamma)^{13}$O reaction may play a role. The curve 1 represents the equilibrium line between the rates for $^{12}$N proton capture and $^{12}$N beta decay. The curve 3 illustrates the same result as determined from Ref. [3]. The curve 2 shows the line of equal strength between the rate of the $^{12}$N radiative proton capture to $^{13}$O and the rate for the inverse process, $^{13}$O photodisintegration. See text for details.

In Fig. 10, we also compare the results of this work with the results of Ref. [3], plotted here by the dashed line. Our revised reaction rate for $^{12}$N$(p,\gamma)^{13}$O implies that it will only compete successfully with $^{12}$N$(\beta^+\nu)^{12}$C and $^{13}$O photodisintegration at much *higher* densities than initially anticipated. Yet, these findings are at variance with the results reported recently in Ref. [23]. The astrophysical $S$ factor found there is similar to the result reported here. The authors, nonetheless, concluded that proton capture on $^{12}$N would compete successfully with $^{12}$N *β*-decay at *lower* stellar densities than the ones found in Ref. [3]. This apparent contradiction may be due to an incorrect evaluation of the reaction rate for $^{12}$N$(p,\gamma)^{13}$O carried out in Ref. [23]. To support this, we note that the equilibrium curve between the $^{12}$N *β*-decay and $^{12}$N proton capture calculated in Ref. [23] is plotted there in the density-temperature diagram below the equivalent curve calculated with the $^{12}$N$(p,\gamma)^{13}$O reaction rate of Ref. [3]. From the inverse proportionality between the density and the reaction rate for the proton radiative capture and the fact that the $S$ factor evaluated in Ref. [23] is much smaller than the $S$ factor evaluated in Ref. [3], it results that the density-temperature diagram as drawn in Ref. [23] simply can not be correct.

## VI. CONCLUSION

The $^{12}$N$(p,\gamma)^{13}$O reaction was investigated indirectly with the ANC method. A ($^{12}$N,$^{13}$O) proton transfer reaction at 12 MeV/nucleon was used to extract the ANC for the virtual synthesis $^{12}$N + $p \rightarrow$ $^{13}$O and calculate from it the corresponding astrophysical $S$ factor.

We determined $C^2_{p_{1/2}}(^{13}O_{g.s.})$ = 2.53 ± 0.30 fm$^{-1}$ and a value $S_{dc}(0)$ = 0.33(4) keV·b for the direct component of the $S$ factor at zero energy. Interference between the direct capture to the ground state of $^{13}$O and the resonant capture through its first excited state leads to a further enhancement yielding $S_{total}(0)$ = 0.42(5) keV·b. This value for the total $S$ factor is two orders of magnitude smaller than the value used previously by Wiescher *et al.* [3]. Consequently our revised reaction rate for $^{12}$N$(p,\gamma)^{13}$O is significantly smaller than the reaction rate initially evaluated by Ref. [3], implying that $^{12}$N$(p,\gamma)^{13}$O will only compete successfully with $^{12}$N$(\beta^+\nu)$ at higher stellar densities than previously anticipated. This may have substantial implications especially for the evolution of massive metal-free stars between 120 and 1200 solar masses that need a metallicity as small as 1 × 10$^{-9}$ to supply nuclear energy generation through the hot CNO cycle for hydrogen burning. Therefore new hydrodynamical calculations including a full and revised nuclear reaction network are needed to validate the scenario proposed by Wiescher *et al.* [3]. Such a scenario would create CNO material by *hot* pp chains and *rap*-processes bypassing the slow $3\alpha$ process, influencing thus the evolution of Population III stars.

## ACKNOWLEDGMENTS

We thank the Accelerator staff of TAMU Cyclotron Institute for providing a good-quality $^{12}$C primary beam. This work was supported in part by the U.S. DOE under grant DE-FG02-93ER40773 and the Robert A. Welch Foundation under grant A-1082.

## References

[1] V. Bromm and R. B. Larson, Annu. Rev. Astron. Astrophys. **42**, 79 (2004).
[2] G. M. Fuller, S. E. Woosley, T. A. Weaver, ApJ **307**, 675 (1986).
[3] M. Wiescher, J. Görres, S. Graff, L. Buchmann, F.-K. Thielemann, ApJ **343**, 352 (1989).
[4] A. Heger, S. E. Woosley, I. Baraffe, T. Abel, in Proc. of the Mpa/Eso/Mpe/Usm Joint Astronom Conference on Lighthouses of the Universe: The Most Luminous Celestial Objects and Their Use for Cosmology, edited by M. Gilfanov *et al.*, (Springer-Verlag, 2002, pp. 369-375).
[5] H. M. Xu, C. A. Gagliardi, R. E. Tribble, A. M. Mukhamedzhanov, N. K. Timofeyuk, Phys. Rev. Lett. **73**, 2027 (1994); A. M. Mukhamedzhanov *et al.*, Phys.Rev. C **56**, 1302 (1997); C. A. Gagliardi *et al.*, Phys. Rev. C **59**, 1149 (1999).
[6] L. Trache, A. Azhari, H. L. Clark, C. A. Glagiardi, Y.-W. Lui, A. M. Mukhamedzhanov, R. E. Tribble, F. Carstoiu, Phys. Rev. C **61**, 024612 (2000); F. Carstoiu, L. Trache, R. E. Tribble, C. A. Gagliardi, Phys. Rev. C **70**, 054610 (2004).
[7] R. E. Tribble, R. H. Burch, C. A. Gagliardi, Nucl. Instr. Meth. Phys. Res. **A285**, 441 (1989).
[8] A. Azhari, V. Burjan, F. Carstoiu, C. A. Gagliardi, V. Kroha, A. M. Mukhamedzhanov, F. M. Nunes, X. Tang, L. Trache, R. E. Tribble, Phys. Rev. C **63**, 055803 (2001).
[9] G. Tabacaru *et al.*, Phys. Rev. C **73**, 025808 (2006).
[10] L. Trache, A. Azhari, H. L. Clark, C. A. Gagliardi, Y.-W. Lui, A. M. Mukhamedzhanov, R. E. Tribble, F. Carstoiu, Phys. Rev. C **58**, 2715 (1998).
[11] X. Tang, A. Azhari, C. A. Gagliardi, A. M. Mukhamedzhanov, F. Pirlepesov, L. Trache, R.E. Tribble, V. Burjan, V. Kroha, F. Carstoiu, Phys. Rev. C **67**, 015804 (2003); X. Tang *et al.*, Phys. Rev. C **69**, 055807 (2004).
[12] J. C. Blackmon *et al.*, Phys. Rev. C **72**, 034606 (2005).
[13] J. P. Jeukenne, A. Lejeune, and C. Mahaux, Phys. Rev. C **16**, 80 (1977).
[14] E. Bauge, J. P. Delaroche, and M. Girod, Phys. Rev. C **58**, 1118 (1998).
[15] J. Raynal, Phys. Rev. C **23**, 2571 (1981).
[16] S. Raman, C. W. Nistor, and P. Tikkanen, At. Data Nucl. Data Tables **78**, 1 (2001).
[17] F. Carstoiu *et al.*, (to be published).
[18] P. Bém *et al.*, Phys. Rev. C **62**, 024320 (2000).
[19] M. Rhoades-Brown, M. H. McFarlane, and S. C. Pieper, Phys. Rev. C **21**, 2417 (1980); **21**, 2436 (1980).
[20] L. Trache, A. Kolomiets, S. Shlomo, K. Heyde, H. Dejbakhsh, C. A. Gagliardi, R. E. Tribble, X. G. Zhou, V. E. Iacob, A. M. Oros, Phys. Rev. C **54**, 2361 (1996).
[21] R. E. Warner *et al.*, Phys. Rev. C **74**, 014605 (2006).
[22] A. M. Mukhamedzhanov, C. A. Gagliardi, and R. E. Tribble, Phys. Rev. C **63**, 024612 (2001).
[23] B. B. Skorodumov *et al.*, Phys. Rev. C **75**, 024607 (R) (2007).
[24] C. Rofs and W. Rodney, *Cauldrons in the Cosmos* (University of Chicago Press, Chicago, 1988), p.158.
[25] C. A. Bertulani, Comput. Phys. Commun. **156**, 123 (2003).
[26] W. A. Fowler, G. R. Caughlan, and B. A. Zimmerman, Ann. Rev. Astr. Ap. **5**, 525 (1967).